\documentclass[twocolumn]{aastex63}

\newcommand{\MS}{\ifmmode{\,}\else\thinspace\fi{\rm M}\ifmmode_{\odot}\else$_{\odot}$\fi}
\newcommand{\LS}{\ifmmode{\,}\else\thinspace\fi{\rm L}\ifmmode_{\odot}\else$_{\odot}$\fi}
\newcommand{\RS}{\ifmmode{\,}\else\thinspace\fi{\rm R}\ifmmode_{\odot}\else$_{\odot}$\fi}
\newcommand{\teff}{\ifmmode T_{\rm eff}\else$T_{\rm eff}$\fi}
\newcommand{\Ke}{\ifmmode{\,}\else\thinspace\fi{\rm K}}

\received{\today}
\revised{\today}
\accepted{\today}

\submitjournal{ApJL}

\shorttitle{OGLE-GAL-ACEP-091 -- The First Known Multi-Mode Anomalous Cepheid}
\shortauthors{Soszy\'nski et al.}

\begin{document}

\title{OGLE-GAL-ACEP-091 -- The First Known Multi-Mode Anomalous Cepheid}

\author{I. Soszy\'nski}
\affiliation{Astronomical Observatory, University of Warsaw, Al.~Ujazdowskie~4, 00-478~Warszawa, Poland}
\email{soszynsk@astrouw.edu.pl}

\author{R. Smolec}
\affiliation{Nicolaus Copernicus Astronomical Center of the Polish Academy of Sciences, ul. Bartycka 18, PL-00-716 Warszawa, Poland}

\author{A. Udalski}
\affiliation{Astronomical Observatory, University of Warsaw, Al.~Ujazdowskie~4, 00-478~Warszawa, Poland}

\author{M. K. Szyma\'nski}
\affiliation{Astronomical Observatory, University of Warsaw, Al.~Ujazdowskie~4, 00-478~Warszawa, Poland}

\author{P. Pietrukowicz}
\affiliation{Astronomical Observatory, University of Warsaw, Al.~Ujazdowskie~4, 00-478~Warszawa, Poland}

\author{D. M. Skowron}
\affiliation{Astronomical Observatory, University of Warsaw, Al.~Ujazdowskie~4, 00-478~Warszawa, Poland}

\author{J. Skowron}
\affiliation{Astronomical Observatory, University of Warsaw, Al.~Ujazdowskie~4, 00-478~Warszawa, Poland}

\author{P. Mr\'oz}
\affiliation{Astronomical Observatory, University of Warsaw, Al.~Ujazdowskie~4, 00-478~Warszawa, Poland}
\affiliation{Division of Physics, Mathematics, and Astronomy, California Institute of Technology, Pasadena, CA 91125, USA}

\author{R. Poleski}
\affiliation{Astronomical Observatory, University of Warsaw, Al.~Ujazdowskie~4, 00-478~Warszawa, Poland}

\author{S. Koz\l{}owski}
\affiliation{Astronomical Observatory, University of Warsaw, Al.~Ujazdowskie~4, 00-478~Warszawa, Poland}

\author{P. Iwanek}
\affiliation{Astronomical Observatory, University of Warsaw, Al.~Ujazdowskie~4, 00-478~Warszawa, Poland}

\author{M. Wrona}
\affiliation{Astronomical Observatory, University of Warsaw, Al.~Ujazdowskie~4, 00-478~Warszawa, Poland}

\author{M. Gromadzki}
\affiliation{Astronomical Observatory, University of Warsaw, Al.~Ujazdowskie~4, 00-478~Warszawa, Poland}

\author{K. Ulaczyk}
\affiliation{Astronomical Observatory, University of Warsaw, Al.~Ujazdowskie~4, 00-478~Warszawa, Poland}
\affiliation{Department of Physics, University of Warwick, Gibbet Hill Road, Coventry, CV4~7AL,~UK}

\author{K. Rybicki}
\affiliation{Astronomical Observatory, University of Warsaw, Al.~Ujazdowskie~4, 00-478~Warszawa, Poland}

\begin{abstract}

Anomalous Cepheids (ACs) are metal-deficient, core-helium-burning pulsating stars with masses in the range 1.2--2.2\MS. Until recently, all known ACs were pure single-mode pulsators. The first candidate for an AC pulsating in more than one radial mode -- OGLE-GAL-ACEP-091 -- was recently identified in the Milky Way based on the photometric database of the Optical Gravitational Lensing Experiment (OGLE) survey. We analyze this object showing that it is actually a triple-mode pulsator. Its position in the Petersen diagram, the light-curve morphology quantified by Fourier coefficients, and absolute magnitudes derived from the Gaia parallax are consistent with the assumption that OGLE-GAL-ACEP-091 is an AC. Our grid of linear pulsation models indicates that OGLE-GAL-ACEP-091 is a 1.8\MS\ star with a metallicity of about ${\rm [Fe/H]}=-0.5$~dex.

\end{abstract}


\section{Introduction}\label{sec:intro}

Among the diversity of variable stars, multi-mode pulsators play a unique role, because two, three or more periods detected in the same star strictly constrain stellar parameters like mass, luminosity, and metal content. \citet{fath1937} and \citet{florja1937} were the first who correctly interpreted strong amplitude variations observed in the light curves of some pulsating stars as a superposition of two or more periodic components. \citet{fath1937,fath1940} discovered three periods in $\delta$~Sct \citep[an object that two decades later was established as a prototype of a new type of pulsating stars;][]{eggen1956}. In turn, \citet{florja1937} detected two periods of a triple-mode pulsator AC And, much later classified by \citet{fernie1994} as an intermediate between $\delta$~Sct stars and classical Cepheids. The first bona fide double-mode (or beat) classical Cepheid, U TrA, was discovered by \citet{oosterhoff1957}. AQ Leo was the first known double-mode RR Lyr variable \citep{jerzykiewicz1977}. Nowadays, hundreds of multi-mode classical Cepheids and RR Lyr stars are known in the Milky Way \citep[e.g.][]{udalski2018,soszynski2019,soszynski2020}.

For years, all type II Cepheids were considered to be pure fundamental-mode pulsators, until \citet{smolec2018} identified the first two cases of beat type II Cepheids in the OGLE Collection of Variable Stars (OCVS). Currently, a total of five objects belonging to this class are known in the Galaxy \citep{udalski2018,soszynski2020}.

\begin{figure*}
\plotone{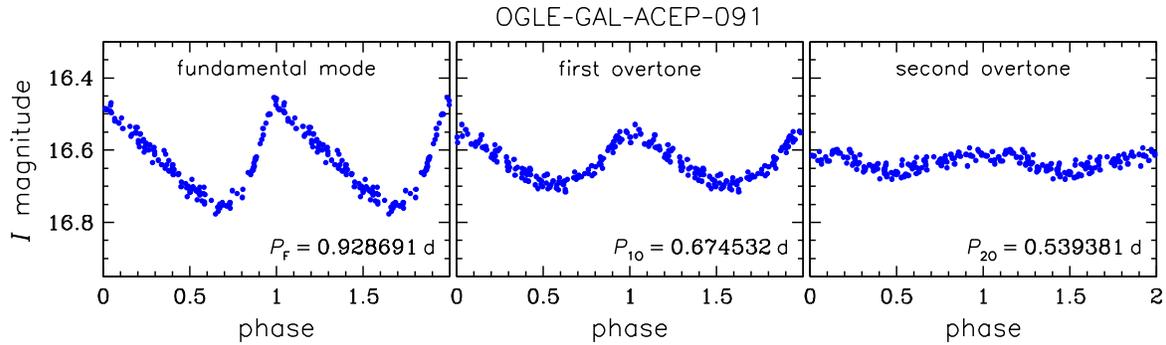}
\caption{Disentangled {\it I}-band light curves of OGLE-GAL-ACEP-091. Left, middle, and right panels show light curves of F, 1O, and 2O modes, respectively, prewhitened for the other modes and for linear combination terms.\label{fig1}}
\end{figure*}

In contrast to other members of the Cepheid family, until recently all known anomalous Cepheids (ACs) were single-mode stars, pulsating exclusively in the fundamental or first-overtone mode. ACs are metal-poor core-helium-burning stars with masses ranging from 1.2 to 2.2\,\MS\ \citep[e.g.][]{bono1997,marconi2004,groenewegen2017a}. The fundamental-mode ACs have pulsation periods in the range of 0.6--2.5~d, while the first-overtone variables oscillate with periods of 0.4--1.2~d. These stars obey period--luminosity (PL) relations located between the relations for classical and type II Cepheids. The evolutionary scenario that leads to ACs is not well understood, but various observational facts indicate that members of this class of pulsating stars are products of mass transfer or component merging in close binary systems \cite[e.g.][]{gautschy2017,iwanek2018}.

The first ACs were discovered by \citet{thackeray1950} in the Sculptor dwarf galaxy. Over the next half century, several dozen ACs have been identified, mostly in nearby dwarf spheroidal galaxies. This number was significantly increased by the discovery of ACs in the Large and Small Magellanic Clouds (LMC and SMC) in the OGLE photometric databases \citep[e.g.][]{soszynski2015}. This homogeneous sample of 268 ACs allowed us to develop a method of distinguishing ACs from other types of classical pulsators based on their light curve shape quantified by the Fourier coefficients $\phi_{21}$ and $\phi_{31}$. Then, we used this method to detect over 100 ACs in the bulge and halo of the Milky Way \citep[e.g.][]{soszynski2015,soszynski2020} -- the first unambiguously identified objects of that type in the Galactic field.

\begin{deluxetable}{rl}
\tablecaption{Parameters of OGLE-GAL-ACEP-091\label{tab:table1}}
\startdata
&\\
Right ascension [J2000.0]:       & 17:38:33.27   \\
Declination [J2000.0]:           & $-$32:53:32.2 \\
Galactic longitude [\degr]:      & 355.839819    \\
Galactic latitude [\degr]:       & $-$0.799623   \\
F-mode period, $P_{\rm F}$ [d]:   & 0.928691     \\
1O-mode period, $P_{\rm 1O}$ [d]: & 0.674532     \\
2O-mode period, $P_{\rm 2O}$ [d]: & 0.539381     \\
Mean {\it I}-band magnitude (OGLE):  & 16.63    \\
Mean {\it J}-band magnitude (VVV):   & 12.74    \\
Mean {\it H}-band magnitude (VVV):   & 11.60    \\
Mean {\it K$_S$}-band magnitude (VVV): & 11.13    \\
Mean [3.6]-band magnitude (Spitzer): & 10.32    \\
Mean [4.5]-band magnitude (Spitzer): & 10.24    \\
Mean [5.8]-band magnitude (Spitzer): & 10.08    \\
Mean [8.0]-band magnitude (Spitzer): & 10.37    \\
\enddata
\end{deluxetable}

This work is devoted to the study of OGLE-GAL-ACEP-091 -- the recently identified candidate for a Galactic AC pulsating simultaneously in three radial modes: fundamental (F), first-overtone (1O) and second-overtone (2O). We provide arguments that this object is indeed the first known multi-mode AC. Basic parameters of OGLE-GAL-ACEP-091 are presented in Table~\ref{tab:table1}.

\section{Observational Data}\label{sec:obs}

A triple-mode pulsating star OGLE-GAL-ACEP-091 was identified by \citet{soszynski2020} in the photometric databases collected by the OGLE Galaxy Variability Survey \citep{udalski2018}. The OGLE project monitors the densest regions of the sky using 1.3-m Warsaw telescope located at Las Campanas Observatory, Chile. The telescope is equipped with a 268 Megapixel mosaic camera composed of 32 CCD chips with a pixel scale of 0.26~arcsec. Details of the instrumental setup, data reduction, and calibration of the OGLE data can be found in \citet{udalski2015}. In our investigation we use the OGLE time-series photometry obtained in the Cousins {\it I}-band filter between March 2017 and March 2020 (in total 117 data points). Typical errors of individual measurements range from 0.01 to 0.02~mag. Figure~\ref{fig1} shows {\it I}-band light curves of the three modes, each one after subtracting the two other frequencies and their linear combinations.

\section{Discussion}\label{sec:dis}

\subsection{Position in the Petersen Diagram} \label{sec:pet}

\begin{figure*}
\centering
\includegraphics[width=13cm]{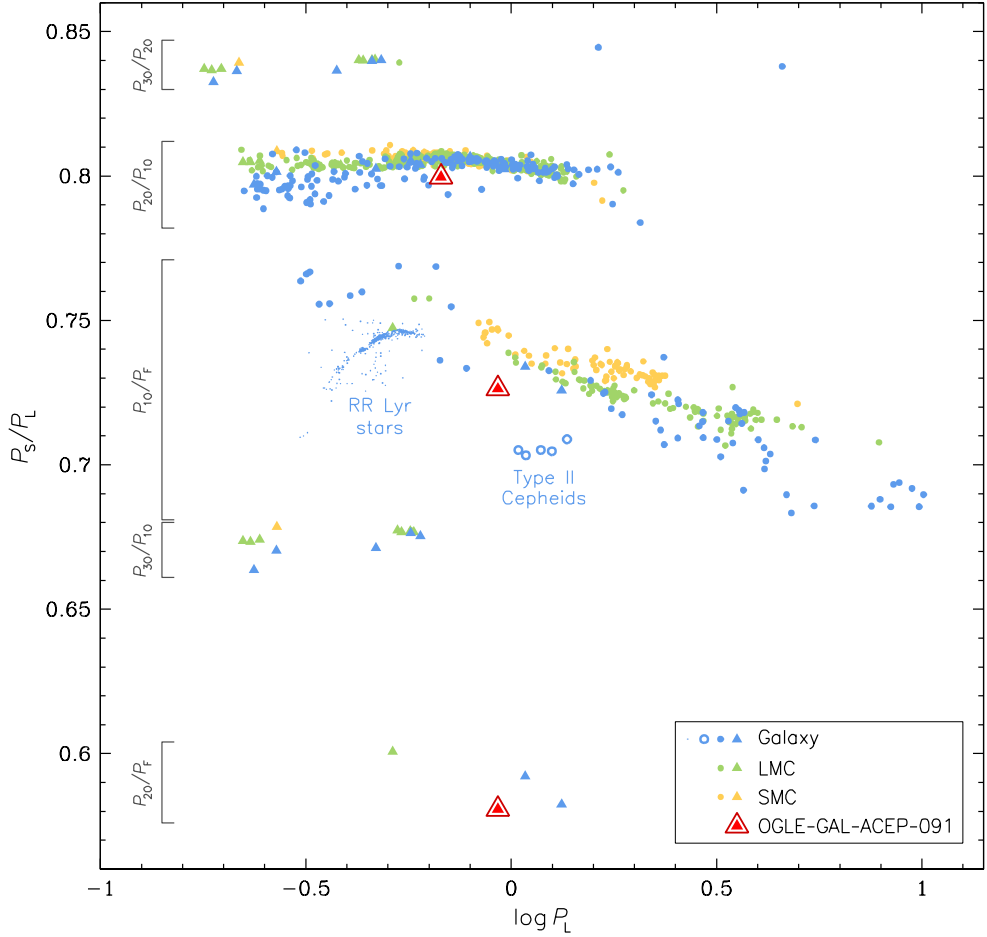}
\caption{Petersen diagram for multi-mode Cepheids and RR Lyr stars in the Milky Way (blue symbols), LMC (green symbols), and SMC (yellow symbols). Filled circles mark double-mode Cepheids, triangles -- triple-mode Cepheids, empty circles -- double-mode type II Cepheids, dots -- RR Lyr (RRd) stars. Period ratios of OGLE-GAL-ACEP-091 are marked by red triangles.\label{fig2}}
\end{figure*}

Figure~\ref{fig2} shows the Petersen diagram (shorter-to-longer period ratio plotted against logarithmic longer period) for multi-mode Cepheids and RR Lyr stars from the OCVS. Blue, green and yellow symbols indicate variables in the Milky Way, LMC, and SMC, respectively. OGLE-GAL-ACEP-091 is marked by three red triangles corresponding to the three period ratios. The positions of OGLE-GAL-ACEP-091 on the Petersen diagram confirms that this star pulsates simultaneously in the F, 1O, and 2O modes. However, it is worth noting that the 1O to F-mode period ratio, $P_{\rm 1O}/P_{\rm F}$, is smaller than the $P_{\rm 1O}/P_{\rm F}$ ratios for classical Cepheids of similar periods. This suggests that OGLE-GAL-ACEP-091 is not a classical Cepheid and belongs to a different class of pulsating stars.

\subsection{Fourier Coefficients}\label{sec:fou}

\begin{figure*}
\centering
\includegraphics[width=13cm]{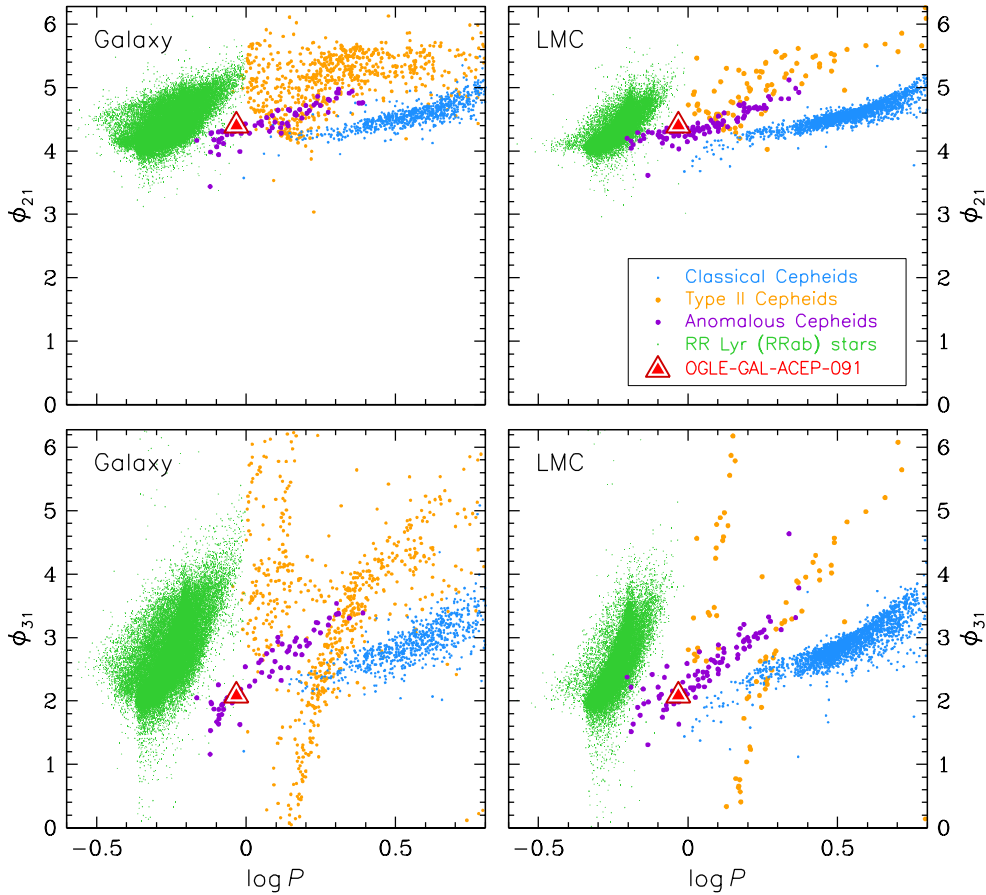}
\caption{Fourier parameters $\phi_{21}$ and $\phi_{31}$ as a function of $\log{P}$ for F-mode classical pulsators in the Galaxy (left panels) and LMC (right panels). Blue points represent classical Cepheids, orange points -- type II Cepheids, green points -- RR Lyr stars, purple points -- single-mode anomalous Cepheids. Parameters of the F-mode light curve of OGLE-GAL-ACEP-091 are marked with red triangles.\label{fig3}}
\end{figure*}

To recognize the variability type of OGLE-GAL-ACEP-091, we fitted its F-mode {\it I}-band light curve (left panel of Figure~\ref{fig1}) with a truncated Fourier series in the following form:
\begin{equation}
I(t)=I_{0}+\sum_{k=1}^{5}A_{k} \cos(\frac{2 \pi k t}{P}+\phi_{k})
\label{eq:fourier}
\end{equation}

Then, we derived the Fourier coefficients $\phi_{21}=\phi_2-2\phi_1$ and $\phi_{31}=\phi_3-3\phi_1$ and plotted them against period in Figure~\ref{fig3} together with the F-mode classical pulsators in the Galaxy and LMC: classical, type II and anomalous Cepheids, and RR Lyr (RRab) stars. As shown by \citet{soszynski2015}, the $\log{P}$--$\phi_{21}$ and $\log{P}$--$\phi_{31}$ diagrams are the most efficient tools for distinguishing ACs from classical Cepheids and RRab stars. The position of OGLE-GAL-ACEP-091 in the period--Fourier coefficients diagrams clearly indicates that it is an AC.

\subsection{Distance}\label{sec:dist}

On the PL plane, ACs populate the region between the PL relations delineated by classical Cepheids and the PL sequences defined by type II Cepheids and RR Lyr stars \citep[e.g.][]{soszynski2015,groenewegen2017b}. The F-mode ACs in the LMC are about 0.7~mag fainter than F-mode classical Cepheids with the same periods. To place OGLE-GAL-ACEP-091 on the absolute PL diagram, we need to know its distance from the Sun. Unfortunately, the Gaia DR2 parallax of this star \citep{gaia2018} is known with the relative uncertainty reaching 40\%. The parallax $0.58\pm0.23$~mas corresponds to a distance of $2.0^{+2.8}_{-0.8}$~kpc \citep{bailer-jones2018}.

This value can be compared with the distances derived from the PL relations for classical Cepheids and ACs. Since OGLE-GAL-ACEP-091 is observed toward the Galactic bulge, close to the Galactic plane, it is affected by heavy interstellar extinction. Therefore, we used the mid-infrared magnitudes obtained by the Spitzer Space Telescope through the GLIMPSE II program \citep{churchwell2009}. Following the procedure presented by \citet{skowron2019}, we used the three-dimensional map of interstellar extinction ``mwdust'' \citep{bovy2016} in conjunction with the infrared PL relations for classical Cepheids \citep{wang2018} and ACs (assumed to be 0.7~mag fainter than classical Cepheids in all filters). The distances to OGLE-GAL-ACEP-091 derived for four Spitzer passbands -- [3.6], [4.5], [5.8], and [8.0] -- are consistent with each other within errors. Assuming that OGLE-GAL-ACEP-091 follows the PL relation for classical Cepheids, its distance from us would be $2.93\pm0.11$~kpc (the mean value from four Spitzer bands). If OGLE-GAL-ACEP-091 obeyed the PL relation for ACs, its distance would be $2.27\pm0.10$~kpc, which better agrees with the parallax from Gaia. This is another argument in favor of the claim that OGLE-GAL-ACEP-091 is an AC.

\subsection{Mode selection in ACs}\label{sec:mode}

Of the several hundred ACs currently known in the Milky Way and nearby galaxies, OGLE-GAL-ACEP-091 is the only case with more than one radial mode simultaneously excited. Why ACs avoid multimodality? It is not easy to answer this question, because the mode selection in pulsating stars is a nonlinear process \citep[see][for a review]{smolec2014}. However, the linear stability calculations may provide some clues to this issue. One may conclude that the double-mode F+1O pulsation is not possible, or is less likely, if the F-mode and 1O-mode instability strips are decoupled in the HR diagram, or have only a little common part.

\begin{figure}
\centering
\includegraphics[width=8.5cm]{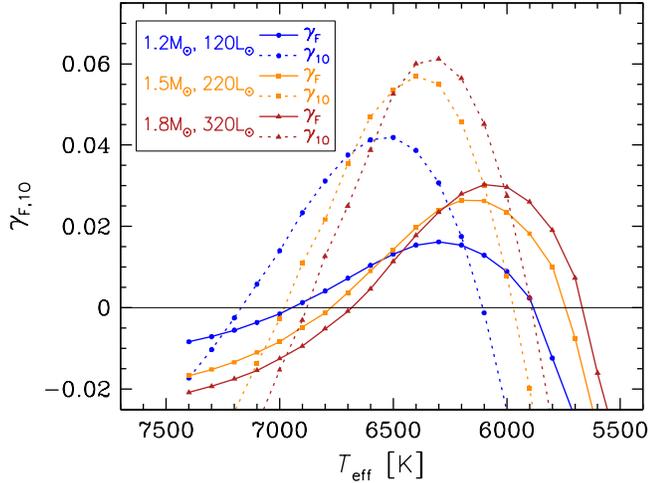}
\caption{Linear growth rates for radial F and 1O modes plotted vs. effective temperature for three sequences of models with metallicity ${\rm [Fe/H]}=-1.5$~dex and constant mass and luminosity, as indicated in the key.\label{fig4}}
\end{figure}

To check this hypothesis, we computed a grid of linear pulsation models with the \cite{smolec2008} convective pulsation code implementing the \cite{kuhfuss1986} model of convection-pulsation coupling. OPAL opacities \citep{opal1996} and \cite{asplund2009} scaled solar mixture are adopted in all calculations. Convective parameters are the same as in set B in \cite{mesa5}. We have considered a range of masses, $1.2-2.2\MS$, a range of luminosities, $100-400\LS$ and a range of metallicities, from metal poor (${\rm [Fe/H]}=-2.5$~dex), to solar (${\rm [Fe/H]}=0.0$~dex). We note that ACs are believed to be metal poor. At fixed $M$/$L$, models were computed with a $100$\,K step in effective temperature, covering the full extent of the instability strip for the F mode and the lowest order overtones.

In Figure~\ref{fig4}, we show the growth rates for the F mode and 1O mode, $\gamma_{\rm F}$ and $\gamma_{\rm 1O}$, respectively, computed for ${\rm [Fe/H]}=-1.5$~dex and for three different combinations of $M$ and $L$, that lead to the pulsation periods in the range covered by ACs ($0.6<P_{\rm F}<2.5$\,d). Growth rates, both $\gamma_{\rm F}$ and $\gamma_{\rm 1O}$, are plotted versus the effective temperature (consecutive models are spaced by 100 K). The plots are qualitatively similar for other metallicities and other $M$/$L$ combinations. The two radial modes are simultaneously unstable for a significant part of the HR diagram, which is similar as for RR~Lyrae stars or for classical Cepheids. Consequently, the lack of the double-mode F+1O ACs cannot be explained at the linear level. Nonlinear calculations are beyond the scope of this paper.

\subsection{Linear modelling of OGLE-GAL-ACEP-091}\label{sec:mode}

\begin{figure}
\includegraphics[width=8.5cm]{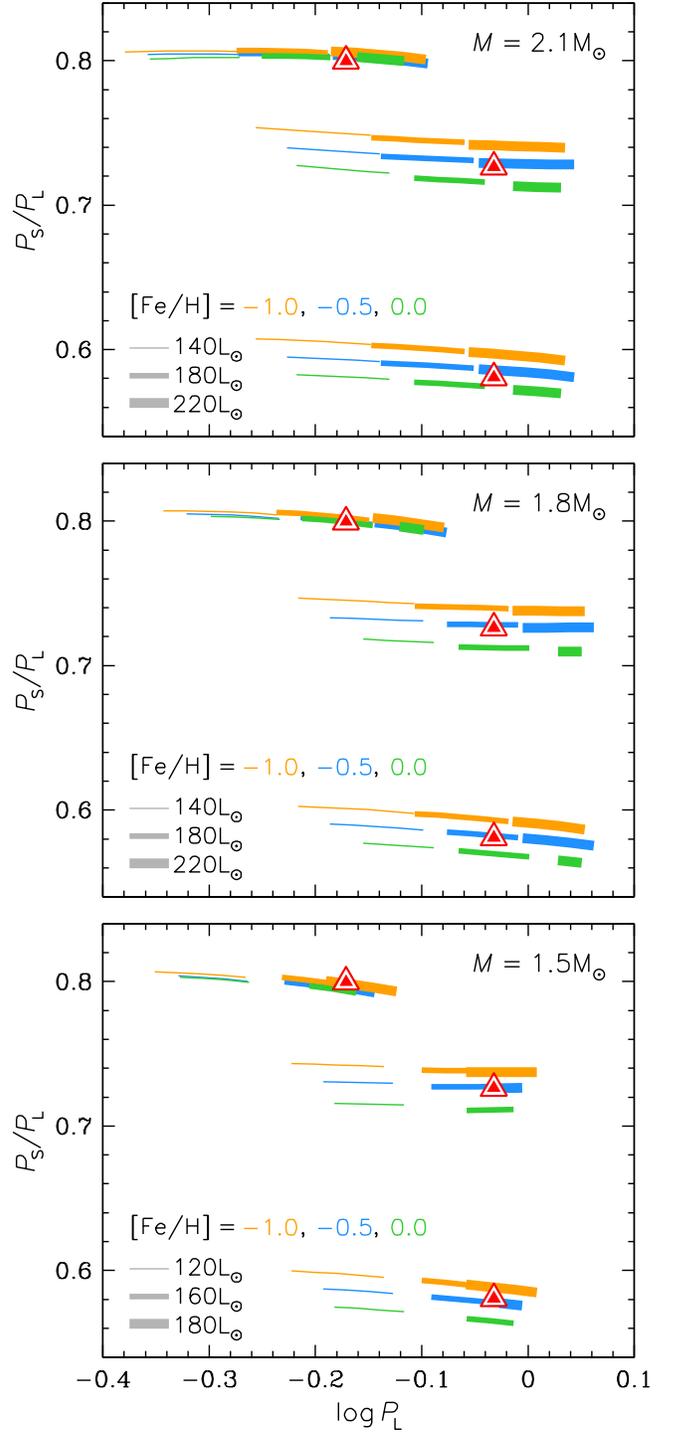}
\caption{Petersen diagrams with linear pulsation models for OGLE-GAL-ACEP-091 (red triangles). The top, middle and bottom panels, correspond to models of $M=2.1, 1.8$ and $1.5$\,\MS, respectively. In each panel, line segments correspond to horizontal model sequences of constant $L$ (value indicated with line thickness) and effective temperature varying in 100\,K steps, in which F, 1O and 2O are simultaneously unstable. Models for three different metallicities are plotted with different colors.\label{fig5}}
\end{figure}

Using the model grid presented in the previous section, we find that a very good match for the periods detected in OGLE-GAL-ACEP-091 can be found only for high metallicity models of relatively large mass and of low luminosity. For lower metallicities, the period ratios are too high as compared with the observations. For lower masses and/or higher luminosities the three modes are not simultaneously unstable. The best matching models are illustrated in Figure~\ref{fig5} in which we show the Petersen diagrams computed for $M=2.1\MS$ (top panel) $M=1.8\MS$ (middle panel) and $M=1.5\MS$ (bottom panel), a range of luminosities plotted with different line thickness, as indicated in each panel, and for three different metallicities, ${\rm [Fe/H]}=-1.0,\,-0.5,\,0.0$~dex,  indicated with different colors. The line segments correspond to models of the just specified parameters in which three radial modes are simultaneously unstable -- a necessary condition for non-resonant triple-mode pulsation, while red triangles represent the three period ratios observed in OGLE-GAL-ACEP-091.

We observe the following: (i) The $P_{\rm 2O}/P_{\rm 1O}$ period ratio shows little sensitivity to ${\rm [Fe/H]}$. It shows some sensitivity to mass -- the models with $M=1.8\MS$ fit the $P_{\rm 2O}/P_{\rm 1O}$ best. (ii) The $P_{\rm 1O}/P_{\rm F}$ and $P_{\rm 2O}/P_{\rm F}$ are both sensitive to metallicity. Models with ${\rm [Fe/H]}=-0.5$~dex match the observed period ratios best. (iii) All period ratios are little sensitive to the value of the luminosity, but with increasing luminosity the domain in which three radial modes are simultaneously unstable shifts to longer periods. (iv) By slightly adjusting the metallicity around ${\rm [Fe/H]}=-0.5$~dex, a satisfactory match to all period ratios can be found for each of the masses considered in the figure. For $M=1.5\MS$ well matching models are found for $L=160\LS$, while for $M=2.1\MS$ the best matching luminosity is $L=220\MS$. 

At the linear level we cannot constrain the physical parameters of OGLE-GAL-ACEP-091 further. We can, however, compare the resulting luminosity with the observed one, estimated using the period-luminosity relation for F-mode ACs published by \cite{groenewegen2017b}. This way we can estimate the luminosity of OGLE-GAL-ACEP-091 to $\approx 110\LS$. It is smaller than the best matching model luminosities given above, but in our opinion is not in conflict with the models. First, the PL relation has some dispersion; second, we cannot exclude a metallicity effect on the PL relation for ACs. While the calibrated relation is based on the metal poor single-mode pulsators, OGLE-GAL-ACEP-091 may be a much more metal rich star, as indicated with its lower period ratios. However, note that for the metallicity range covered in the Magellanic Clouds, \cite{groenewegen2017b} claim no metallicity effect on the PL relation. Still, the comparison with the luminosity based on the PL relation favors lower masses and luminosities of the models ($M=1.5\MS$, $L=160\LS$).

The fact that satisfactory models can be found assuming masses and luminosities in the range typically considered for ACs is another confirmation that the classification of OGLE-GAL-ACEP-091 as an AC is correct. The linear pulsation models  point that its metallicity is around ${\rm [Fe/H]}=-0.5$~dex, higher than typically assumed for ACs, which may be a hint to mode selection problem discussed in the previous section. Higher metallicity may favor double-mode or multi-mode  pulsations, while typical ACs are metal poor and hence typically single-periodic.

The inferred high metallicity of OGLE-GAL-ACEP-091, corresponding to $Z\approx 0.004$, is in conflict with a single-star evolution scenario of ACs, as metal-rich ($Z\gtrsim 0.001$) intermediate-mass stars do not cross the instability strip \citep[see][for a review]{bono2016}. Large population of ACs in the relatively metal-rich LMC motivated theoretical studies to understand formation mechanisms of ACs \citep[see e.g.][]{fiorentino2012,gautschy2017}. In the latter study, interactions in the binary system, including merger events, were considered. It was shown that models with significantly higher metallicity, up to $Z=0.008$, may pass through the instability strip. While we do not have observational evidence in support of binary evolution of OGLE-GAL-ACEP-091, such scenario cannot be excluded. Undoubtedly, the discovery of a larger number of multi-mode ACs may provide useful constraints on their physical parameters and evolutionary scenarios.

\section{Conclusions}
We showed that OGLE-GAL-ACEP-091 is a triple-mode AC -- the first such object known in the Universe. The Fourier coefficients $\phi_{21}$ and $\phi_{31}$ derived for the F-mode light curve perfectly match to F-mode ACs found by the OGLE team in the Milky Way and LMC. Also the Gaia DR2 parallax suggests that OGLE-GAL-ACEP-091 belongs to the AC family, although a more accurate parallax measurement is necessary to dispel any doubts.

Our linear pulsation models confirm that indeed the position of OGLE-GAL-ACEP-091 in the Petersen diagram is consistent with a simultaneous excitation of three lowest order radial modes in models with masses and luminosities in the range typically considered for ACs. While ACs are considered metal deficient stars, models indicate that the metallicity of the triple-mode pulsator is relatively high, around ${\rm [Fe/H]}=-0.5$~dex. It may suggest that high metallicity is a factor that favors multi-mode pulsation in ACs.

\acknowledgments
We thank Dr. Giuseppe Bono, the referee of this paper, for his valuable comments on this work. This paper has been supported by the National Science Centre, Poland, grant MAESTRO no. 2016/22/A/ST9/ 00009. The OGLE project has received funding from the Polish National Science Centre grant MAESTRO no. 2014/14/A/ST9/00121. RS was supported by the National Science Center, Poland, Sonata BIS project 2018/30/E/ST9/00598. This work is based in part on observations made with the Spitzer Space Telescope, which was operated by the Jet Propulsion Laboratory, California Institute of Technology under a contract with NASA.

\end{document}